\RequirePackage{fix-cm}
\documentclass[twocolumn]{svjour3}          
\smartqed  
\usepackage{graphicx}
\usepackage{mathptmx}      
\usepackage{latexsym}
\usepackage[usenames, dvipsnames]{color}

\journalname{Biological Cybernetics}

\begin{document}

\title{Making decisions in the dark basement of the brain: A look back at the GPR model of action selection and the basal ganglia}
	

\titlerunning{The GPR model: a retrospective}        

\author{Mark D. Humphries         \and
        Kevin Gurney 
}


\institute{Mark D. Humphries \at
              University of Nottingham \\
              \email{mark.humphries@nottingham.ac.uk}           
			\and
           Kevin Gurney \at
              University of Sheffield \\
              \email{k.gurney@sheffield.ac.uk} 
}

\date{Received: date / Accepted: date}

\maketitle

\begin{abstract} 
How does your brain decide what you will do next? Over the past few decades compelling evidence has emerged that the basal ganglia, a collection of nuclei in the fore- and mid-brain of all vertebrates, are vital to action selection. Gurney, Prescott, and Redgrave published an influential computational account of this idea in Biological Cybernetics in 2001. Here we take a look back at this pair of papers, outlining the ``GPR'' model contained therein, the context of that model's development, and the influence it has had over the past twenty years. Tracing its lineage into models and theories still emerging now, we are encouraged that the GPR model is that rare thing, a computational model of a brain circuit whose advances were directly built on by others.

\end{abstract}

\section{Introduction}
\label{intro}
On the first day of my PhD in October 1998 I (MDH) was handed a thick, 51-page report, held between covers of pale blue card on which were printed the ominous words ``Analysis and simulation of a model of the basal ganglia,'' and instructed by my supervisor, one Kevin Gurney, to read all the contents therein. Intimidating as it was for a first year PhD student fresh from their undergraduate studies, this behemoth of a technical report would become the foundation of a pair of classic papers published together in Biological Cybernetics in 2001, both with the same author line of Gurney, Prescott and Redgrave: ``A computational model of action selection in the basal ganglia I: A new functional anatomy'' \cite{Gurney2001} and ``A computational model of action selection in the basal ganglia II: Analysis and simulation of behaviour'' \cite{Gurney2001a}. 

Here we take a look back at this pair of papers, at their context, their influence, and what the future may hold for the model of the brain they contain. In reference to their author line, and in keeping with the common name they have acquired over the last 20 years, we refer to the model as the GPR model throughout.
	
\section{Why the basal ganglia at all?}
\begin{figure*}[t!]
  \includegraphics[width=\textwidth]{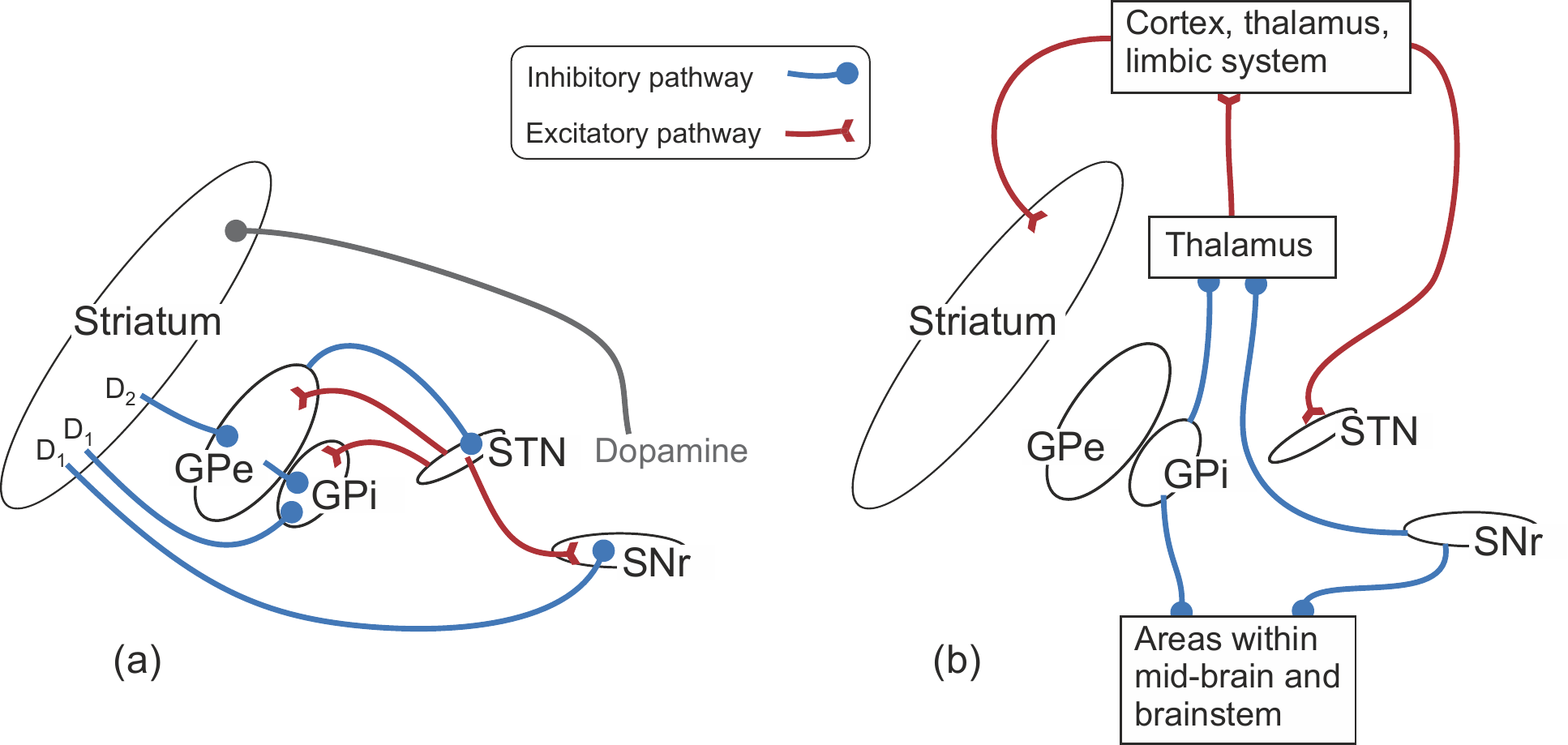}
\caption{\textbf{Anatomy of the basal ganglia.} (a) Nuclei of the basal ganglia nuclei and their internal connections. Nuclei positions and sizes correspond to their approximate relative locations and sizes in a sagittal section of the rodent brain. Dopamine is released into the striatum by inputs from the midbrain, and striatal neurons express either the D1 or D2 type of dopamine receptor. (b) External connections of the basal ganglia to the rest of the brain. Rectangular boxes indicate regions outside the basal ganglia. Neurons of the globus pallidus pars interna (GPi) and substantia nigra pars reticulata (SNr) receive similar inputs and are the main source of projections outside of the basal ganglia, and so are collectively termed the ``output nuclei'' here. GPe: globus pallidus pars externa; STN: subthalamic nucleus.}
\label{fig:bg_anatomy}       
\end{figure*}

The basal ganglia comprise the massive striatum, sitting underneath the cortex in much of the forebrain, and a group of much smaller deep-lying nuclei (Figure \ref{fig:bg_anatomy}). Their functional role has long perplexed us: Kinnear-Wilson called them the dark basement of the brain as far back as the 1920s \cite{Graybiel2000}, a feeling still evoked in many of its researchers now, not least because they are seemingly involved in so many disorders. It is from the striatum that dopamine is lost when midbrain dopaminergic neurons die in Parkinson's disease, leading to the classic clinical signs of akinesia, rigidity, and tremor. The death of the principle, projection neurons of the striatum leads the appearance of chorea, the unpredictable, uncontrolled limb movements in Huntington's disease. So it had long been thought that, whatever they do, it involved movement in some way \cite{Denny-Brown1962}.  

By the late 1990s a consensus was emerging around the idea that the basal ganglia are crucial to the selection of movements, but not their initiation \cite{Mink1993,Marsden1994a,Mink1996}. This idea beautifully explained how the basal ganglia could be seemingly so central to such a wide-range of movement disorders, how stimulating neurons within them could immediately evoke movement, and yet damage to the basal ganglia did not prevent movement from happening. What it left open was \emph{how}: how exactly did the basal ganglia select movements? 

At the time, sketched ideas for this centred on the concept of disinhibition \cite{Deniau1985,Chevalier1990}. Neurons of the basal ganglia's output nuclei are persistently active, at around 60 spikes per second in primates, and GABAergic, so constantly inhibiting every neuron they target. The disinhibition concept proposed that turning off this persistently inhibitory output signals the selection of the motor program encoded by the target neurons.

At about the same time, there emerged influential conceptual models of the wiring between these nuclei. One model proposed two pathways from the striatum that converged on the output nuclei of the basal ganglia. The striatal neurons of the so-called ``direct'' pathway sent their axons directly to the neurons of the output nuclei; the striatal neurons of the ``indirect'' pathway sent their axons to the globus pallidus pars externa (GPe in Fig \ref{fig:bg_anatomy}), whose neurons in turn project to the output nuclei \cite{Albin1989,Alexander1990,DeLong1990}. Another model proposed that the basal ganglia were organised topographically, so connections between nuclei formed parallel loops \cite{Alexander1986,Alexander1990}. Moreover, it was becoming clear that connections from the tiny subthalamic nucleus, the only source of excitatory glutamate within the basal ganglia, to the output nuclei were a further key player \cite{Parent1995}. Unclear was how to link these conceptual models of wiring to the conceptual models of selection. 

Into this milieu were launched what was intended to be a triptych of papers. The first, bylined Redgrave, Prescott, and Gurney \cite{Redgrave1999}, argued how the idea of movement selection can be understood as a special case of the formal problem of action selection, and showed how the basal ganglia were seemingly ideally placed to solve that problem. Not least that these parallel loops were, at a fine scale, the substrate for representing competing actions. The second, bylined Prescott, Redgrave and Gurney \cite{Prescott1999} put this hypothesis into the context of the rest of the brain, arguing that the brain contains many-layered control of movement, stretching from the deepest brainstem to the cortex, and the basal ganglia has privileged access to many of these layers, positioning it at a key locus for controlling actions. Together, these papers made a compelling case for a more nuanced account for what the basal ganglia do: they select actions.

The third planned paper, which transpired to be the pair of papers that are our subject here, was the \emph{how}: the model that synthesised all the above conceptual ideas into a single quantitative computational account of how the basal ganglia implement action selection. And it answered two crucial questions missing from all the words and the box-and-arrow diagrams \cite{Albin2001}: how does it resolve competition between actions; and how does it switch actions?

\section{The GPR model}

\begin{figure*}[t!]
	\includegraphics{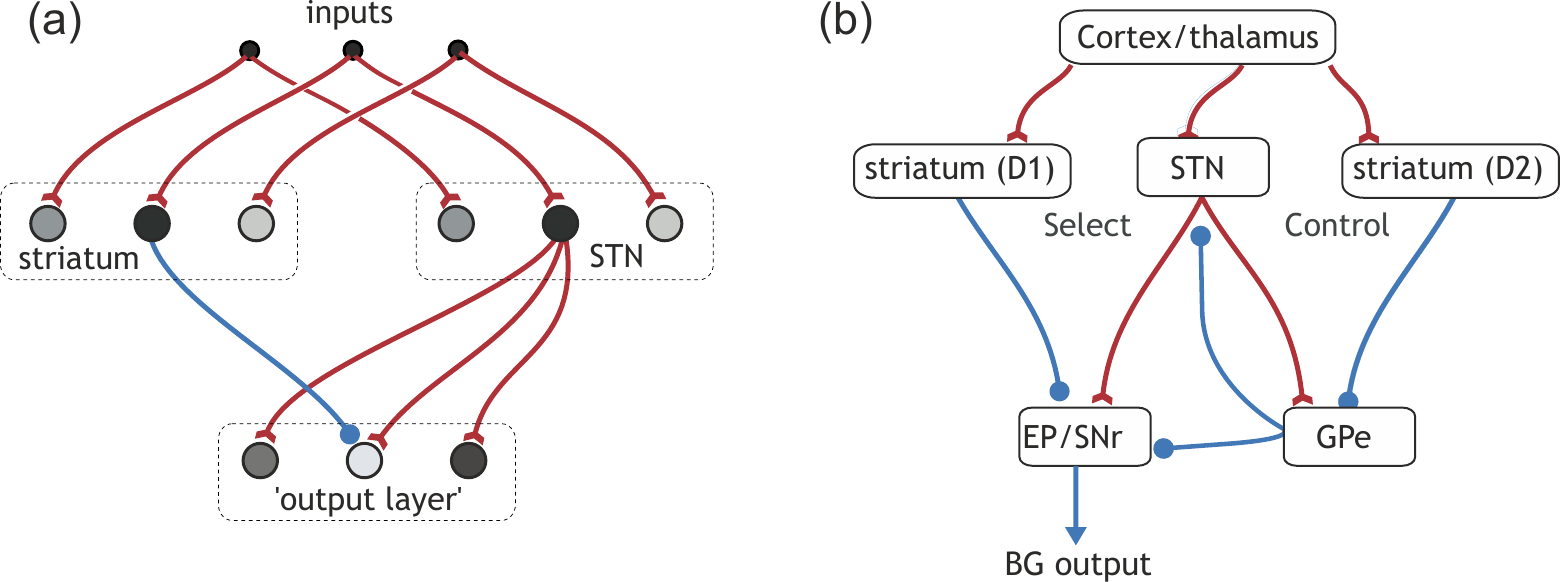}
	\caption{\textbf{Functional anatomy of the basal ganglia}. As laid out in Gurney et al \cite{Gurney2001}. (a) The anatomy of how selection works. Populations or ``channels'' in each nucleus represent actions (circles). In this reading of the anatomy, populations of D1-expressing striatal neurons send inhibitory projections to their corresponding populations in the output nuclei; whereas populations in the subthalamic nucleus (STN) send diffuse excitatory projections to all populations in the output nuclei (we show only full connections of the central channel for clarity). Grey-scale fills indicate an example of relative levels of activity in each ``channel'', with dark-to-light encoding high-to-low activity. In this example, the combination of focussed inhibition and diffuse excitation arriving at the output layer results in the central channel having little activity and the others more activity, indicating clean selection of the central channel's action. (b) Full functional anatomy, with interpretations of the functional role of the anatomically defined pathways.}
	\label{fig:bg_fcn_anatomy}       
\end{figure*}

The first paper \cite{Gurney2001} tackled the job of synthesising the above ideas and data into a single coherent whole, into the circuit for selection defined by the wiring of the basal ganglia; it also did the important job of formally defining what selection means. Figure \ref{fig:bg_fcn_anatomy} shows the resulting ``functional anatomy'' of the basal ganglia, so called because it is anatomy read in the light of a functional hypothesis. This formalised how actions could be represented within the basal ganglia by parallel groups of neurons -- ``channels'' in the paper's parlance. In this scheme, each channel through the basal ganglia represents an action, and the neurons' activity within a channel represents the salience of that action. The action with the highest salience ought to win the competition between the currently available actions, and be selected. Selection was defined using the concept of disinhibition: the selected channel of the output nucleus was the one whose activity sufficiently reduced for its inhibitory influence to be removed.   

In the second paper \cite{Gurney2001a} was laid out the formal mathematical model built on that anatomy, its analytical solutions, and simulations that illustrated their key points. The model was constructed at a population level, each channel within a given nucleus modelled as the ensemble activity of its neurons. A key element of the model was the modulation by dopamine of the striatal neuron's activity, building on recent data showing striatal neurons with dopamine D1 receptors formed the direct pathway and those with D2 receptors the indirect pathway. Critically, the paper synthesised data that suggested activating these receptors has opposite effects on neural activity, with D1 activation enhancing activity and D2 activation depressing activity. 

This model of dopamine has proved remarkably resilient, and remains one of the key contributions of the GPR model. Of the further immediate contributions of this pair of papers, we highlight three.

First is that the new functional anatomy provided an influential bauplan for theories of the basal ganglia. As we discuss below, this template has formed the basis for numerous models of the basal ganglia from us and others, and brought into sharp focus mysteries of the basal ganglia, such as the function of the striatum's internal wiring, some of which have yet to be resolved.

Second is the explanation for how the basal ganglia resolves the competition between its inputs. As laid out in the functional anatomy, the output nuclei receive an odd combination of focused inhibition from striatum and putative diffuse excitation from the subthalamic nucleus (Figure \ref{fig:bg_fcn_anatomy}a). By combining the channel idea with this anatomy, the paper proposed the basal ganglia act as an off-centre, on-surround network: competition between competing actions is realised by the balance of the focussed inhibition and broad excitation received by the neurons of each channel in the output nuclei. The winning channel -- and hence action -- would thus be the one whose balance of inputs tipped most strongly in favour of the inhibition, and so had the most reduced activity. 

This mechanism also explains how actions are switched between. When a more salient action becomes available, the model showed that the same off-centre, on-surround design naturally handles this: the new input further increases the output of the subthalamic nucleus, in turn increasing the activity of all the output channels, thus cancelling selection of the current action; while at the same time the corresponding new, stronger inhibitory signal from striatum inhibits its target population in the output nuclei.

A third contribution is the idea of capacity scaling, that the basal ganglia are able to automatically scale the activity of the output nuclei to cope with the number of competing actions. The problem facing the basal ganglia is that the subthalamic nucleus output is both excitatory and diffuse, so increasing its total input will increase activity across the output nuclei (Figure \ref{fig:bg_fcn_anatomy}a). Consequently, the more actions there are competing, the more input the subthalamic nucleus receives, the harder it is to select a single action by turning off one channel of the output nucleus.  

But as shown in Figure \ref{fig:bg_fcn_anatomy}b, the subthalamic nucleus and globus pallidus form a negative feedback loop. Analysis in the second paper \cite{Gurney2001a} showed that this was sufficient to automatically scale the activity in the subthalamic nucleus: the higher it is driven by its inputs, the more strongly it drives activity in the globus pallidus, which in turn inhibits activity in the subthalamic nucleus. Which all means that the subthalamic nucleus activity is automatically scaled down as the number of its inputs increases. Consequently, the activity of the output nuclei does not grow towards saturation as the number of competing actions increases -- their capacity is scaled. No matter how many extra active inputs join the competition for selection, this feedback loop guarantees that the output activity will remain responsive to its striatal input, and thus allow selection to continue.  

\section{Influence of the model}
The most immediate influence of the GPR model was the fecund research programme it sparked in our group \cite{Gurney2004,Redgrave2021}. We embedded the model in the wider thalamocortical loop within which the basal ganglia sit, and explored how that influenced its capacity for action selection \cite{Humphries2002a}. We created a large-scale spiking neuron version, now with hundreds of individual neurons in each nucleus, to show that the action selection hypothesis was compatible with both the detailed dynamics of individual nuclei and with a wide range of electrophysiological data \cite{Humphries2006c}. With the same spiking model we went on to show how deep brain stimulation of the subthalamic nucleus, an effective treatment for the cardinal motor signs of Parkinson's disease, can elicit a mixture of excitatory and inhibitory responses in the output nuclei \cite{Humphries2012a}. Embodying the model in a mobile robot showed how smooth behavioral sequences naturally emerge from the changing saliences of each action -- but also revealed where our understanding of action selection was lacking \cite{Prescott2006}. In a comprehensive review of the ventral basal ganglia, the part that has the nucleus accumbens as its input structure, we showed that the circuit originating from the core of the accumbens also conformed to the GPR model, thus casting its known roles in navigation and learning as a function of selecting appropriate responses \cite{Humphries2010,Khamassi2012}. We even subjected the poor thing to the Stroop task \cite{Stafford2004,Stafford2007}. This body of work culminated in a complete theory of how the plasticity of the connections from the cortex to the striatum, controlled by dopamine signals that convey prediction errors, drives the learning and extinction of action selection via the basal ganglia \cite{Gurney2015}.

The GPR model found an immediate home in the work of Agnes Guillot and her colleagues, as part of the Psikharpax project to construct an artificial rat. Again embodied in a variety of mobile robots, Guillot's team explored its ability to control foraging \cite{Girard2003}, integrate its action selection with models for navigation \cite{Girard2005}, and be the ``Actor'' in Actor-Critic models of reinforcement learning \cite{Khammasi2005}. Benoit Girard has continued this vein of work into new areas, including a dynamical systems analysis of the model \cite{Girard2008}, and a major piece of work to establish a version that more fully captured the basal ganglia circuit in primates \cite{Lienard2014} -- a full spiking-neuron version of that primate model has recently followed \cite{Girard2020}

A version of the GPR model of the basal ganglia also sits at the heart of SPAUN \cite{Eliasmith2012}, Chris Eliasmith and colleagues' 2.5 million spiking neuron model of a brain architecture for cognitive tasks. Naturally, this basal ganglia model handles the action selection process in all tasks SPAUN is trained to do. Critical to its place in SPAUN though is their addition of plasticity between the cortical-striatal connections \cite{Stewart2012}, essential for the model to relearn associations between the states of the world and the appropriate actions in them for each new task SPAUN faces. 

Rafal Bogacz took the GPR model in a different direction. He noticed that the functional architecture laid out in the first paper \cite{Gurney2001} could potentially implement the multi-sequential probability ratio test (the MSPRT), a Bayesian algorithm for making a decision based on the accumulating evidence for multiple options. In this version of the model, each channel represented an option, the input to striatum represents the momentary evidence to be accumulated, and the basal ganglia output at each moment in time represents the (negative log) of the conditional probability of each option given the evidence so far -- the negative log so that the decision is made when one of the outputs falls below a threshold, again making use of the disinhibition concept. The resulting paper \cite{Bogacz2007}, co-authored with one of us (KG), was perhaps the first to draw formal links between the decision making literature and action selection ideas. The role of the basal ganglia in decision-making is now a burgeoning research field, with deep development of the theory that the basal ganglia implement an optimal algorithm for decision-making \cite{Zhang2010,Lepora2012,Bogacz2016a,Caballero2018}, combined with compelling experimental evidence that the striatum plays a key, causal role in decision making \cite{Ding2010,Ding2012,Yartsev2018a}. 

The GPR model's lineage can be traced to many recent models that have used its ideas as a springboard for a deep exploration of the basal ganglia. These include Fountas and Shanahan's \cite{Fountas2017} spiking model that explored how oscillations in the input to the basal ganglia altered their consequent output, Dunovan and colleagues' \cite{Dunovan2019} spiking model that linked the dynamics of the basal ganglia to the behavioural parameters of decision making, including that the rate of evidence accumulation was defined by the difference in activity between channels of the ``direct'' pathway, and Lindahl and Kotaleski's large-scale spiking model \cite{Lindahl2016} that exhaustively explored what the connections within the basal ganglia contributed to both their dynamical repertoire and their ability to perform action selection. And showing it's still going strong nearly 20 years on, Gilbertson and Steele made use of a variant of the GPR model in their proposal that the combination of dopamine's immediate and long-term effects in the striatum can allow the basal ganglia to optimally solve the exploration-exploitation trade-off in action selection  \cite{Gilbertson2020}.

\section{The future}
Any good model of a specific neural circuit also sets out a research programme for the future, by showing what we don't know, and what we need to know. One such unknown was the contribution of the internal circuitry of the striatum. At the time, we knew of three interneuron types \cite{Kawaguchi1993}, and of the anatomical evidence for connections made between the projection neurons of the striatum \cite{Wilson1980}, but little else; now the number of interneurons types has at least doubled, and we know the microcircuit of the striatum in some detail from both anatomical and electrophysiological data \cite{Tepper2018}. But as the GPR model illustrated, the basal ganglia did not need this striatal microcircuit to perform action selection - so what was it for? This gap in our knowledge has driven a programme of computational work on the striatum's microcircuit by us \cite{Humphries2009,Humphries2010b,Tomkins2014} and others \cite{Yim2011,Spreizer2017}, culminating in a full-scale model of biophysical striatum from Hellgren-Kotaleski and colleagues \cite{Hjorth2020}.   

Our knowledge of the basal ganglia is ever-evolving. In a further example, whereas the globus pallidus is treated as a single entity in the GPR model, we now know it contains at least two distinct populations, the proto-pallidal and arky-pallidal, with different inputs, targets, and dynamics (for review see \cite{Gittis2014}). Moreover, the feedback connections from the globus pallidus to striatum, only hinted at by data available during the model's development in the late 90s \cite{Walker1989,Rajakumar1994}, are now a well-studied pathway that originate from the arkypallidal population \cite{Bevan1998,Mallet2012,Corbit2016}. The venerable GPR model has been updated to integrate these new data too \cite{Suryanarayana2019}.

But as our knowledge of the basal ganglia continues to evolve, the reader may have the nagging question of whether it is time to finally abandon the GPR model, to move on to a new view of how the basal ganglia function. And even if we do not make any explicit decision to abandon a model, we are still left with the computational equivalent of the Ship of Theseus paradox: that if we keep updating and altering the model with each new advance in our knowledge of the basal ganglia brought by experimental data, then what remains of the original? 

Perhaps that is the wrong view of what a model is for. Rather, like all good models, we know the GPR model was flawed at the outset: its test was not whether it was `right', but whether it was useful. And as we hope we have demonstrated here, the GPR model has been decidedly useful to us and many others over the 20 years since its publication, and so shall exist in some form for as long as that may continue. 


%

\begin{acknowledgements}
We thank Peter Redgrave and Tony Prescott for a reading a draft of this retrospective.
\end{acknowledgements}

%
\section*{Conflict of interest}

The authors declare that they have no conflict of interest.

\bibliographystyle{spmpsci}      
\bibliography{GPR_BiolCyber}   


\end{document}